\documentclass{INTERSPEECH2023}
\usepackage{marvosym}
\usepackage{amsmath}

\interspeechcameraready

\usepackage{enumitem}
\usepackage{bm}
\usepackage{caption}
\usepackage{multirow}
\usepackage{subfigure}
\usepackage{booktabs}
\title{MAVD: The First Open Large-Scale Mandarin Audio-Visual Dataset with Depth Information}
\name{Jianrong Wang$^{1}$ \qquad Yuchen Huo$^{2}$ \qquad Li Liu$^{3}$$^{(\textrm{\Letter})}$ \thanks{This work was supported by the National Natural Science Foundation of
China (No. 61977049) and National Natural Science Foundation of China (No. 62101351).}\qquad Tianyi Xu$^{1}$ \qquad Qi Li$^{4}$ \qquad Sen Li$^{1}$}
\address{
  $^{1}$College of Intelligence and Computing, Tianjin University, Tianjin, China \newline
  $^{2}$Tianjin International Engineering Institute, Tianjin University, Tianjin, China \newline
  $^{3}$The Hong Kong University of Science and Technology (Guangzhou), Guangzhou, China \newline
  $^{4}$School of Electrical and Information Engineering, Tianjin University, Tianjin, China
  }
\email{ avrillliu@hkust-gz.edu.cn}

\begin{document}

\maketitle

\begin{abstract}
Audio-visual speech recognition (AVSR) gains increasing attention from researchers as an important part of human-computer interaction. However, the existing available Mandarin audio-visual datasets are limited and lack the depth information. To address this issue, this work establishes the MAVD, a new large-scale Mandarin multimodal corpus comprising 12,484 utterances spoken by 64 native Chinese speakers. To ensure the dataset covers diverse real-world scenarios, a pipeline for cleaning and filtering the raw text material has been developed to create a well-balanced reading material. In particular, the latest data acquisition device of Microsoft, Azure Kinect is used to capture depth information in addition to the traditional audio signals and RGB images during data acquisition. We also provide a baseline experiment, which could be used to evaluate the effectiveness of the dataset. The dataset and code will be released at https://github.com/SpringHuo/MAVD.
\end{abstract}

\section{Introduction}
Speech recognition is a key component of the human-computer interaction (HCI), serving as an indispensable bridge that connects human-machine communications. Traditionally, early speech recognition systems are capable of recognizing input audio signals and converting them into corresponding text units \cite{first-ref}, which can assist in HCI \cite{sec-ref}, language learning \cite{language}, and aiding people with physical disabilities \cite{disable,liu2019pilot}. Previous research on speech recognition has obtained significant achievements. Currently, it was reported that automatic speech recognition (ASR) systems performed almost as accurately as humans in relatively quiet laboratory environments \cite{panda}.

However, audio-based speech recognition systems still cannot achieve satisfactory performance in noisy environments. This is because human speech recognition involves the acquisition of information from multiple modalities and an extensive analysis for a relatively accurate result \cite{hear1976,liu2020re}. To address this issue, visual signals that remain unaffected by noise were presented to provide supplementary information to the speech signal \cite{input}. The research results indicated that visual information is advantageous to ASR, particularly in the presence of noisy or unavailable audio \cite{avsravaliable}.

Moreover, the quality and quantity of data used can greatly affect the performance and accuracy of deep neural networks (DNNs). In addition to traditional datasets containing audio and color images, researchers wanted additional information to complement the visual feature. Proposed solutions included using multi-view capture devices and stereo cameras, which have been shown to improve the audio-visual speech recognition (AVSR) model based on experimental evidence \cite{avaliablemulti, multiview,liu2018automatic,liu2018visual}. However, the availability of Mandarin open-source audio-visual datasets appears to be quite limited as it is time and labor-consuming. In conclusion, existing datasets still suffer from the following challenges: 

\begin{itemize}[nosep,left=2em]

\item  There is a shortage of Mandarin reading materials that are relevant to daily life and gathered from a wide range of domains. 

\item The annotations of the previous datasets are unitary, and thus cannot be directly adapted to various scenarios (e.g., sentence-level or phoneme-level).

\item Due to the lack of the Mandarin audio-visual dataset with depth information, the speaker's head or body movements, and the noise caused by the recording environment may have a negative impact on feature extraction.

 \end{itemize}

To solve the above-mentioned challenges, we establish a large-scale $\textbf{M}$andarin $\textbf{A}$udio-$\textbf{V}$isual dataset with $\textbf{D}$epth information, named MAVD. The corpus contains 12,484 utterances spoken by 64 speakers. The distribution of speakers covers 24 different provinces of China, and all speakers are asked to record in Mandarin. The corpus of raw text comprises content from a variety of popular open-source media in China (e.g. Weibo, People's Daily, etc.) to better fit the words used in daily life. All raw text is obtained through an automated pipeline and then these texts are filtered manually to avoid some immoral or violent sentences. Azure Kinect is utilized to collect data, which includes depth images, color images, and audio. Information about the speaker is given along with multimodal data, including gender, age, hometown, etc. We annotate the audio at the phoneme-level and the \textit{pinyin} sequences, tone sequences, part-of-speech combinations, di-phone, and tri-phone sequences of each sentence are obtained by using the automatic language dictionary-based rule to enlarge the annotations.

By making our corpus open-source and completely free for research use, we intend to contribute to the Mandarin audio-visual speech recognition community. As far as we know, this is the first open large-scale Mandarin audio-visual dataset with depth information. 

\begin{table*}[]
\centering
\caption{Overview of recent audio-visual datasets which are related with ours. For the different modalities, A for audio, V for Visual (RGB) and D for Depth information.}
\label{Table 1}
\begin{tabular}{ccccccc}
\toprule
\textbf{DataSet} & \textbf{Language} & \textbf{Modalities} & \textbf{Speakers} & \textbf{Samples} & \textbf{Recoding Angle}                & \textbf{Scenario} \\ \hline
AV-Letters [15]        & English           & A, V                 & 10                & 780              & Front                                  & Indoor            \\
GRID [18]              & English           & A, V                 & 34                & 1,000             & Front                                  & Indoor            \\
CUAVE [27]            & English           & A, V                & 36                & 7,000             & 0$^\circ$, 90$^\circ$, -90$^\circ$     & Indoor            \\
GAMVA [28]            & Japanese          & A, V                 & 20                & 500              & Eleven different angles                 & Indoor        \\   
BAVCD [30]           & English           & A, V, D               & 21                & 6,700             & Front                                  & Indoor            \\
RGB-D [31]               & English           & A, V, D              & 53                 & 1,060             & Front                                  & Indoor           \\
CAVSR [19]           & Mandarin          & A, V                 & 20                & 3,120             & Front                                  & Indoor            \\
LRW-1000 [23]        & Mandarin          & A, V                 & 2,000              & 718,018           & Front                                  & Wild              \\ \hline
MAVD (ours)                & Mandarin          & A, V, D               & 64                & 12,484            & Front                                  & Indoor            \\ \bottomrule
\end{tabular}
\end{table*}

\section{Related work}

In the literature, there were some audio-visual datasets in multiple languages. Table \ref{Table 1} outlines some of the most prominent and widely utilized datasets.
\subsection{Audio-Visual Datasets}
For traditional audio-visual datasets, early AVSR corpora usually only asked speakers to read some simple and short phrases, such as numbers and letters. The AV-Letters \cite{AVLetter} was one of the most used datasets for training the alphabetic AVSR system. It consisted of 780 utterances spoken by 10 subjects. Every English letter had been repeated 3 times by each speaker. In addition, their team released AV-Letters-2 \cite{AVLetter-2}, which was smaller in size but has been upgraded in terms of recording conditions (primarily improved video quality). Like DAVID \cite{DAVID}, GRID \cite{GRID} collected 1,000 sentences spoken by 34 speakers. Its sentences were simple, syntactically identical phrases. And those phrases consisted of some English words and numbers that are commonly used in daily life. In particular, GRID was recorded in a quiet and noise-free professional studio. These studies demonstrated that improving the recording quality may be beneficial for multimodal speech recognition systems. 

CAVSR was the earliest Chinese audio-visual corpus in Mandarin \cite{cavsr}. This corpus was based on isolated syllables and did not consider continuous pronunciation cases. HIT Bi-CAVDatabase was a dataset mainly for lipreading tasks \cite{hit2,liu2017automatic}, and its subsequent release of the utterance-level continuous lipreading dataset HIT-AVDB II \cite{hit}. In addition, there was a large-scale lip reading dataset called LRW-1000 \cite{LRW-1000}, which contained 718,018 samples from more than 2,000 individual speakers. This was a “speak in the wild” (SITW) dataset \cite{SITW} with most of the data collected from dialogue-based television broadcasts. However, the differences in data sources, including shooting environments and equipment, may cause databases to become more challenging to work with compared to data collected under controlled laboratory conditions \cite{voxceleb,zhang2022webuav}. 

In realistic scenarios, the speaker does not always face the camera but may speak from any angle with complicated phrases. Therefore, it was a common way to set up multi-view devices to capture data from different angles. CUAVE \cite{CUAVE}, containing 7,000 utterances spoken by 36 speakers, with the material of connected or isolated digits. In particular, this dataset had 3 perspectives from the front (0$^\circ$) and two sides (90$^\circ$ and -90$^\circ$). GAMVA \cite{GAMVA} contained 20 Japanese male subjects, each reading 25 sentences of common Japanese phrases. Specifically, GAMVA used 12 cameras to capture information from all possible angles (including pitch and elevation angles). 
\subsection{Audio-Visual Datasets with Depth Information}
Using a stereo camera to record depth information is a more common way to eliminate errors caused by different lighting conditions, as well as unconscious head movements of the speaker. WAPUSK-20 \cite{wapusk20} consisted of 20 subjects reading sentences with the same format as GRID using a Bumblebee stereo-camera capture. Designing and building a 3D-AVSR dataset is always a difficult task due to the complex data synchronization and the expensive cost. This has led researchers to focus on Kinect, which is a more efficient and cheaper device. BAVCD \cite{bavcd} used a first-generation Kinect to capture depth images. This dataset included 6700 utterances read by 15 English-speaking subjects and 6 Greek-speaking subjects. And then in its subsequent work, a new database with multiple views captured the conversation between two subjects. RGB-D \cite{rgbd} used a second-generation Kinect as a 3D information acquisition device to collect data from 20 speakers each reading 20 common English phrases. And through the Kinect, rich 3D data were released, including facial contours, depth images, head orientation, etc. Compared with a large number of multimodal datasets in English, however, available open-source mandarin AVSR databases with depth information are very scarce. 

Therefore, by taking the advantage of each of the above datasets, we use the latest generation of Microsoft Kinect, Azure Kinect, to capture this dataset containing audio, color images, and depth images in a professional soundproof studio.

\section{Our Mandarin Audio-Visual Dataset with Depth Information}

\begin{table*}[]
\centering
\caption{Number of sentences from different data sources, sentence types, and the percentage of the total number of sentences.}
\label{Table 2}
\begin{tabular}{cccc}
\toprule
\multicolumn{2}{c}{\textbf{Material Categorization}} & \textbf{\#Sentences} & \textbf{Percentage (\%)} \\ \hline
\multirow{4}{*}{\textbf{Data Source}} & Social Media          & 6,229 & 49.90 \\
                                      & News                  & 3,135 & 25.11 \\ 
                                      & Book/Novel            & 2,524 & 20.22 \\
                                      & Poetry                & 596  & 4.77  \\ \hline
\multirow{3}{*}{\textbf{Sentence Types}}       & Statement sentences   & 9,573 & 76.68 \\
                                      & Question sentences    & 1,714 & 13.73 \\
                                      & Exclamation sentences & 1,197 & 9.59  \\ \bottomrule
\end{tabular}
\end{table*}

\begin{figure}
\centering 
\includegraphics[height=0.624cm, width = 8cm]{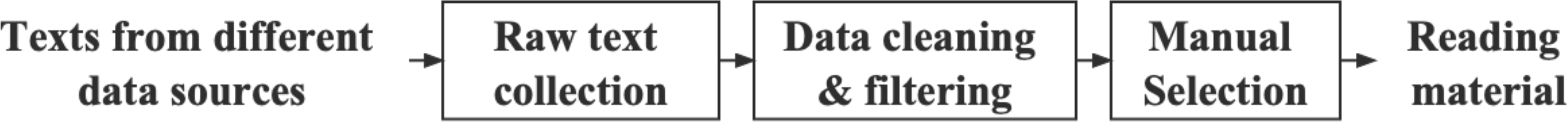}
\caption{Pipeline of reading material acquisition}
\label{fig-pipeline}
\end{figure}

\subsection{Reading Material}
\subsubsection{Text Acquisition Pipeline}
The text acquisition pipeline consists of three parts, raw text collection, data cleaning, and data filtering. The entire acquisition process is shown in Figure \ref{fig-pipeline}. For raw text collection, to ensure the performance of the dataset, it is essential to select raw text from authentic and widely circulated sources with high impact, relevance to daily life, and coverage of a wide range of domains. Based on these principles, we have made a selection of material from various topics such as social media, news, novels, and poetry.

After this, we acquire a large number of raw texts which need to be cleaned first. The data cleaning process includes several steps such as sentence segmentation based on punctuation marks like question marks, semicolons, periods, and exclamation points. Special symbols such as colons, square brackets, English letters, and sensitive words are filtered out. Roman and Arabic numerals are converted to Mandarin expressions, and sentences that are too long or too short are filtered. This process ensures that the raw texts are clean and ready for use in the after accurate analysis.

After data cleaning, we utilize an algorithm to calculate the score of each sentence and select data by sorting the scores. The algorithm combines some metrics commonly found in Mandarin corpora, such as di-phones and tri-phones coverage, tone combination, etc. To be specific, our algorithm weighs the frequency of each metric, attributing a lower score to those with more frequent occurrences. The final score of the sentence is derived from the sum of the factor scores of each sentence. In particular, the types of sentence is taken into account, as the speaker usually shows some natural emotion when reading exclamatory or interrogative sentences. 
\subsubsection{Manual Selection}
After this, all the remaining texts are manually filtered to get the final required reading material. In this process, sentences that cannot be recognized by the machine for sensitive words are removed by manual screening, including sentences with uncommon names, immoral or containing extremely negative messages, etc. The distribution of material categories and the quantities used and the percentage of the total number of sentences are described in Table \ref{Table 2}.

\begin{figure}
\centering 
\includegraphics[height=4.7cm,width=8cm]{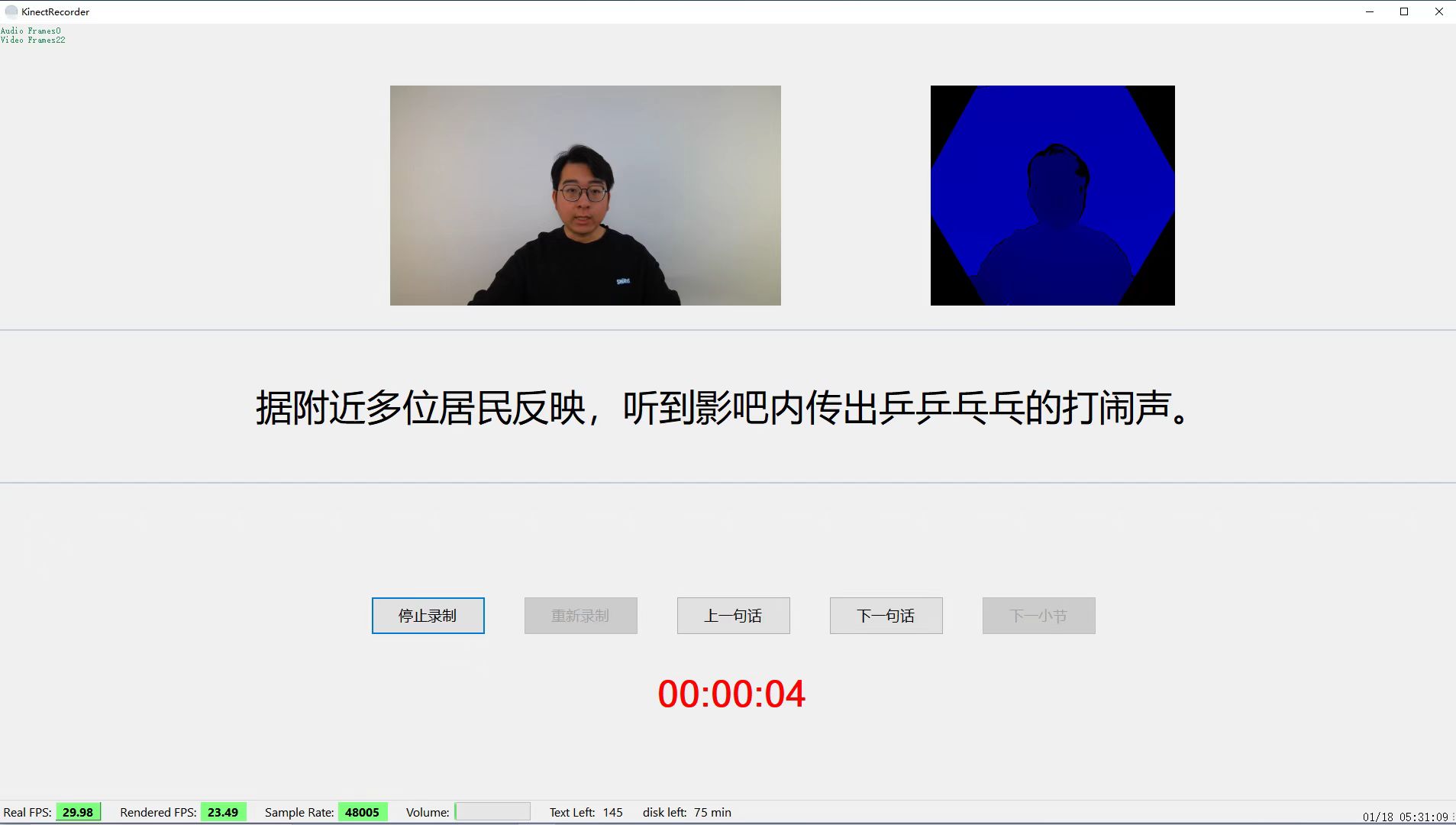}
\caption{The main acquisition interface of the C\#-based data acquisition system, which simultaneously acquires RGB color images, depth images and audio signals.}
\label{fig1}
\end{figure}

\subsection{Speakers}
This dataset consists of 64 speakers, comprising 35 males and 29 females, whose ages range from 19 to 27, with an average age of 23 years. Our criteria for speaker selection are as follows. Firstly, the speaker must not have any constant speech or hearing impairment. Secondly, the gender ratio should be close to 1:1. Finally, the speaker should cover most of the Mandarin-speaking areas as much as possible. The division of speakers according to gender and birthplace is presented in Table \ref{table-gender}.

For research purposes, speaker information including age, gender, hometown, and the presence or absence of an accent was included in our dataset. All speakers signed consent and agreed with the public use of recorded data for speech recognition related research.

\begin{figure}
\centering 
\includegraphics[height=4.7cm,width=8cm]{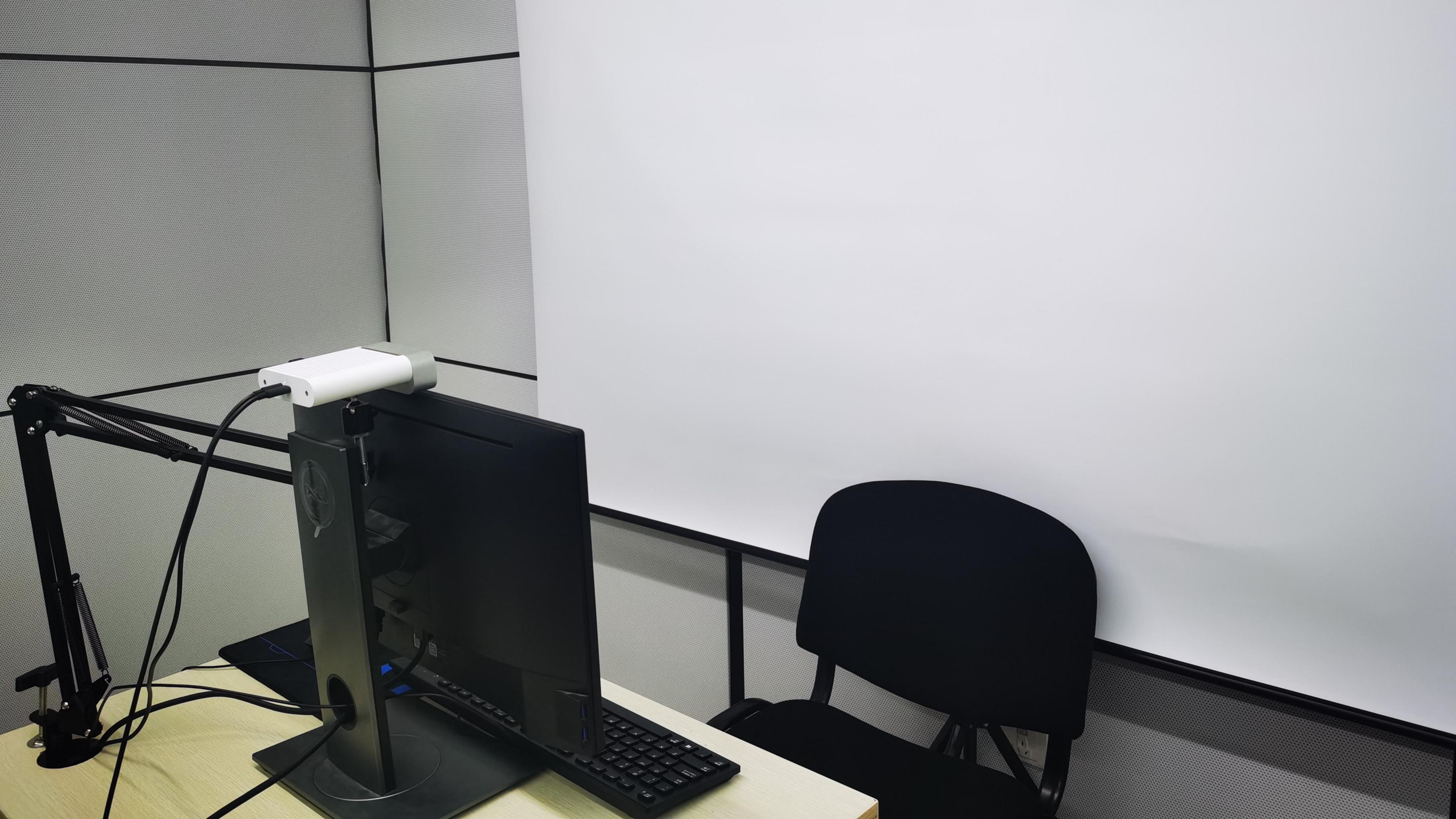}
\caption{Recording environment}
\label{fig2}
\end{figure}

\subsection{Recording Setup}
\subsubsection{Recording Preparation}
As the primary data collection device, Azure Kinect is specially designed for researchers or developers. In the data acquisition, the captured video is consisting of a 30 fps RGBA image stream of 1920$\times$1080 pixels and a 30 fps depth image stream of 640$\times$576 pixels of the narrow field of view (NFoV) unboxed modal. Also, the microphone array of Kinect was used as the audio device.

To capture video and audio effects, we have developed a related capture system based on C\#. The system consists of two parts, the information acquisition page, and the data recording page as shown in Figure \ref{fig1}. On the information acquisition page, the age, gender, hometown, and whether the speaker has an accent are collected. In the formal recording, the speaker reads 200 sentences of text randomly selected from the reading material.

\subsubsection{Environment Setting}
The data recording process takes place in a professional soundproof studio that effectively eliminates sound echoes. The acquisition equipment used consisted of a monitor, the Kinect placed on top of the monitor via a stand, as well as a mainframe running the acquisition system. A white curtain is set up as a background and a fixed speaker position (0.6m-0.7m) to ensure that the depth images are captured correctly. The specific environmental setup is shown in Figure \ref{fig2}.

\begin{table}[]
\centering
\caption{Gender and regional distribution of speakers.}
\label{table-gender}
\begin{tabular}{cccc}
\toprule
\textbf{Gender}         & \textbf{Hometown Area} & \textbf{\#People} & \textbf{Total}      \\ \hline
\multirow{2}{*}{Male}   & North China            & 22                & \multirow{2}{*}{35} \\ 
                        & South China            & 13                &                     \\ \hline
\multirow{2}{*}{Female} & North China            & 23                & \multirow{2}{*}{29} \\ 
                        & South China            & 6                 &                     \\ \bottomrule
\end{tabular}
\end{table}

\subsection{Data Annotation}
First, we use the open-source dictionary generation model grapheme to phoneme (G2P)\footnote{https://github.com/cmusphinx/g2p-seq2seq} to generate a dictionary of text-to-phoneme sequences. After this, a companion TextGrid file with recorded phoneme durations are generated for all captured audio data using the speech-forced alignment tool Montreal Forced Aligner (MFA)\footnote{https://mfa-models.readthedocs.io/en/latest/}.

\begin{figure}[h]
\centering  
\subfigure[RGB image]{   
\begin{minipage}{3.5cm}
\centering    
\includegraphics[height=3.86cm,width=3.5cm]{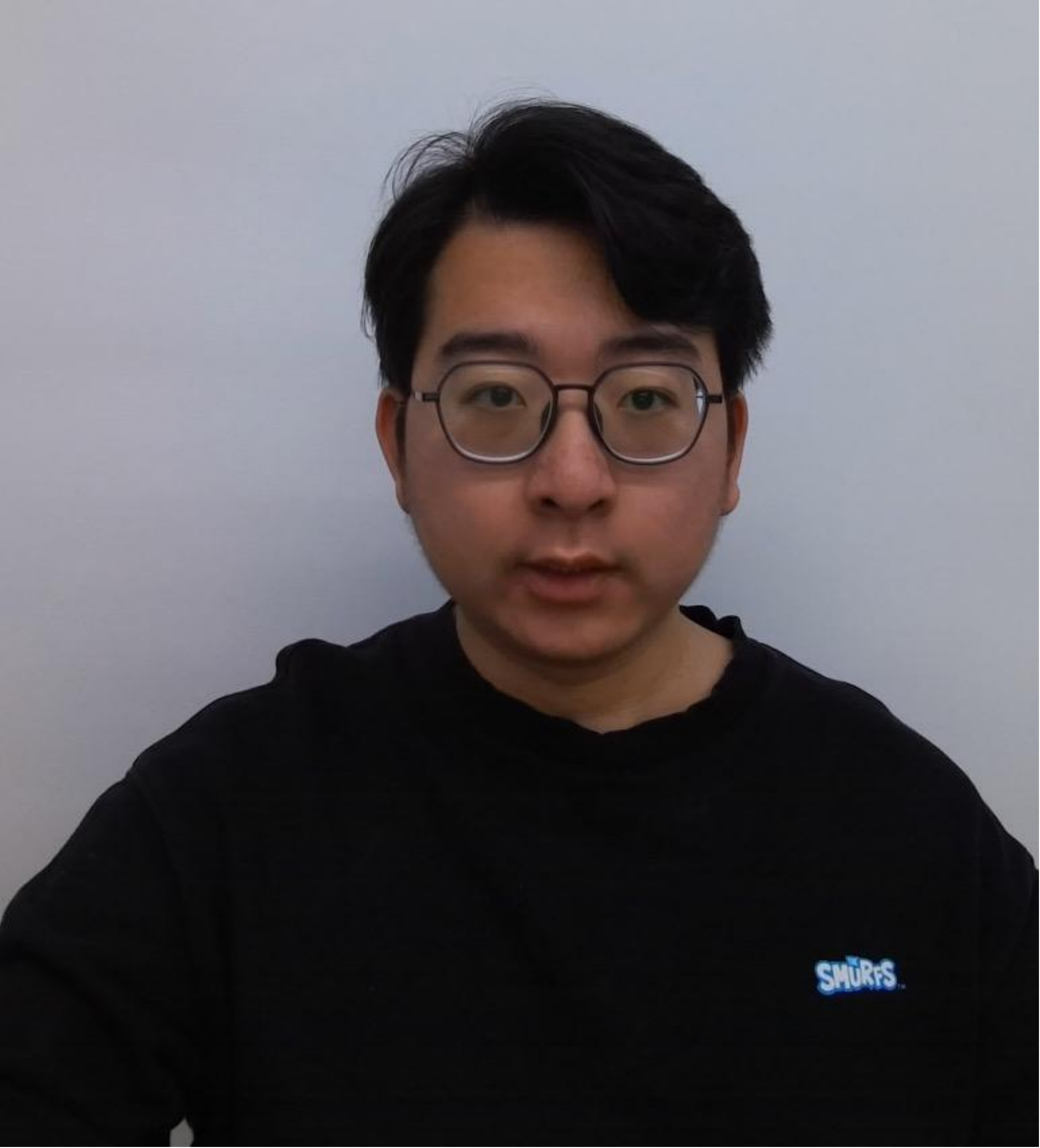}  
\end{minipage}
}
\subfigure[Depth image]{ 
\label{depth image}
\begin{minipage}{3.5cm}
\centering    
\includegraphics[height=3.86cm,width=3.5cm]{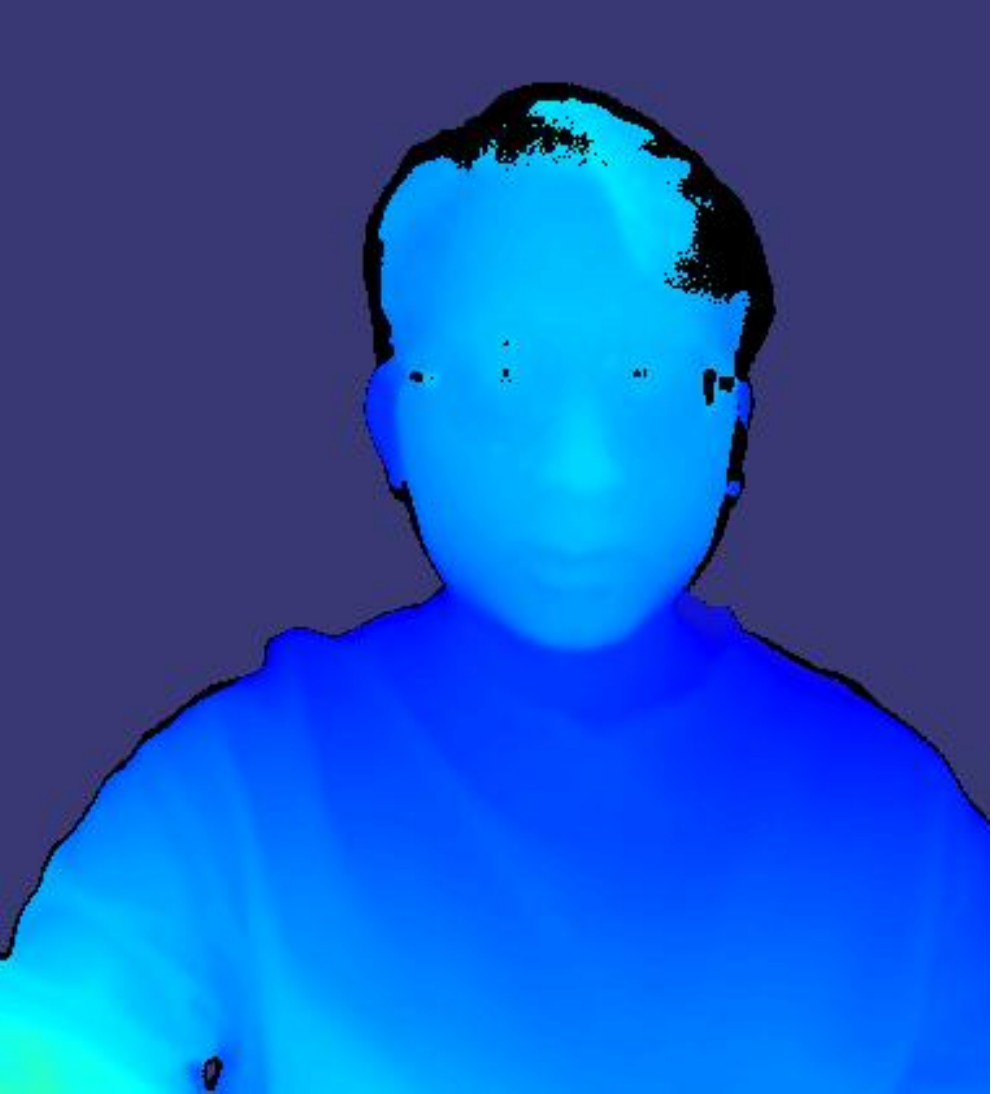}
\end{minipage}
}
\caption{Visual information captured by the Kinect.}    
\label{fig-display}    
\end{figure}

Since the mkv file with RGB image stream, depth image stream, and IR image stream is captured during the data recording process, the image frames need to be extracted and the images need to be named using their timestamps to align with the audio. In particular, in Figure \ref{depth image}, a Look Up Table (LUT) is used to transform the depth image for a more intuitive presentation.

\section{Baseline Experiment}
In this section, we conducted several experiments to assess the efficacy of various modalities in our dataset for speech recognition tasks, by employing a model based on the open-source Wav2Vec 2.0 \cite{wav2vec} and ResNet-18 \cite{resnet}. 
\subsection{Data Pre-processing}
For audio signals, we resample the captured 48kHz audio data to 16kHz, 16-bit, mono-channel. For visual signals, the video is fixed at 25 frames per second to synchronize with the audio and also to avoid the risk of possible frame loss during recording. In addition, the lips are selected as the ROI to extract visual features. By using the pre-trained face landmark detection model in Dlib to detect face landmarks \cite{dlib}, the lip part of each image is cropped. Since the speaker is required not to move significantly during the recording, the lips are positioned every half second (15 frames) to speed up the interception. Finally, both the depth images and the color images, are resized to 64$\times$64 pixels.
\subsection{Baseline Model}
We employ the Mandarin-Wav2Vec2\footnote{https://github.com/kehanlu/Mandarin-Wav2Vec2} open-source model, which was pre-trained on AISHELL-2 \cite{aishell-2} and evaluated on AISHELL-1 \cite{AISHELL-1}, as the acoustic model to extract audio features. For visual signals, The ResNet-18, which was pre-trained on ImageNet, is used as an image recognition model to extract features of both depth and color images.

The audio features and visual features are then input into a Bi-directional Long Short-Term Memory (BiLSTM) network \cite{BILSTM} and collocate the output features. And then as an input to Connectionist Temporal Classification (CTC) decoder to transcribe the text \cite{ctc}.

\begin{table}[]
\centering
\caption{The CER for different input modals on test set.}
\label{table-result}
\begin{tabular}{cc}
\toprule
\textbf{Features}        & \textbf{CER (\%)} \\ \hline
Audio-Only   & 11.53   \\
Audio-Visual (RGB only) & 8.96    \\
Audio-Visual (RGB + Depth)  & 8.78    \\ \bottomrule
\end{tabular}
\end{table}

\subsection{Result and Analysis}
We divided the training, validation, and test set by random selection in the ratio of 8:1:1, under the condition of ensuring gender-balanced. We trained the Wav2Vec-based multimodal on the train set of our data with three different settings by controlling the combination of image features and audio features: (\romannumeral1) audio-only, (\romannumeral2) audio-visual (RGB), and (\romannumeral3) audio-visual (RGB + Depth). The character error rate (CER) is used as an evaluation metric, which is calculated by adding the number of substituted (S), inserted (I), and deleted (D) characters together and then dividing them by the total number of characters (N): $CER = \frac{S + I + D}{N} \times 100.$
 As shown in Table \ref{table-result}, the performance of the audio-only speech recognition is not good mainly due to challenging reading material, and the error rate decreases when combining color and depth images features, which also represents that our proposed corpus is effective in the AVSR task. Importantly, the use of depth information can also help the effectiveness of the Mandarin AVSR task.
\section{Conclusions}
In the work, we present a new multi-modal audio-visual corpus for Mandarin named MAVD, including about 12k  sentences read by 64 speakers. Our corpus focuses on contributing to the Mandarin audio-visual corpus community. As far as we know, this is the first open large-scale Mandarin audio-visual dataset with depth information. In the future, we will continue to expand this work such as using several devices to form a multi-views corpus, working with more recording environments, etc.

\bibliographystyle{IEEEtran}
\bibliography{mybib}

\begin{thebibliography}{10}
\providecommand{\url}[1]{#1}
\csname url@samestyle\endcsname
\providecommand{\newblock}{\relax}
\providecommand{\bibinfo}[2]{#2}
\providecommand{\BIBentrySTDinterwordspacing}{\spaceskip=0pt\relax}
\providecommand{\BIBentryALTinterwordstretchfactor}{4}
\providecommand{\BIBentryALTinterwordspacing}{\spaceskip=\fontdimen2\font plus
\BIBentryALTinterwordstretchfactor\fontdimen3\font minus
  \fontdimen4\font\relax}
\providecommand{\BIBforeignlanguage}[2]{{%
\expandafter\ifx\csname l@#1\endcsname\relax
\typeout{** WARNING: IEEEtran.bst: No hyphenation pattern has been}%
\typeout{** loaded for the language `#1'. Using the pattern for}%
\typeout{** the default language instead.}%
\else
\language=\csname l@#1\endcsname
\fi
#2}}
\providecommand{\BIBdecl}{\relax}
\BIBdecl

\bibitem{first-ref}
T.~Sakai and S.~Doshita, ``{The Automatic Speech Recognition System for
  Conversational Sound},'' \emph{IEEE Transactions on Electronic Computers},
  vol. EC-12, no.~6, pp. 835--846, 1963.

\bibitem{sec-ref}
P.~Sanderson, ``{Cognitive work analysis and the analysis, design, and
  evaluation of human-computer interactive systems},'' in \emph{Proceedings
  1998 Australasian Computer Human Interaction Conference. OzCHI'98 (Cat.
  No.98EX234)}, 1998, pp. 220--227.

\bibitem{language}
E.~M. Golonka, A.~R. Bowles, V.~M. Frank, D.~L. Richardson, and S.~Freynik,
  ``{Technologies for foreign language learning: A review of technology types
  and their effectiveness},'' \emph{Computer assisted language learning},
  vol.~27, no.~1, pp. 70--105, 2014.

\bibitem{disable}
M.~H. Raskind and E.~L. Higgins, ``{Speaking to read: The effects of speech
  recognition technology on the reading and spelling performance of children
  with learning disabilities},'' \emph{Annals of Dyslexia}, vol.~49, pp.
  251--281, 1999.

\bibitem{liu2019pilot}
L.~Liu and G.~Feng, ``A pilot study on mandarin chinese cued speech,''
  \emph{American Annals of the Deaf}, vol. 164, no.~4, p. 496–518, 2019.

\bibitem{panda}
S.~P. Panda, ``{Automated speech recognition system in advancement of
  human-computer interaction},'' in \emph{2017 ICCMC}, 2017, pp. 302--306.

\bibitem{hear1976}
\BIBentryALTinterwordspacing
H.~MCGURK and J.~MACDONALD, ``{Hearing lips and seeing voices},''
  \emph{Nature}, vol. 264, no. 5588, pp. 746--748, Dec. 1976. [Online].
  Available: \url{https://doi.org/10.1038/264746a0}
\BIBentrySTDinterwordspacing

\bibitem{liu2020re}
L.~Liu, G.~Feng, D.~Beautemps, and X.-P. Zhang, ``Re-synchronization using the
  hand preceding model for multi-modal fusion in automatic continuous cued
  speech recognition,'' \emph{IEEE Transactions on Multimedia}, vol.~23, p.
  292–305, 2020.

\bibitem{input}
C.~Ding and D.~Tao, ``{Robust Face Recognition via Multimodal Deep Face
  Representation},'' \emph{IEEE Transactions on Multimedia}, vol.~17, no.~11,
  pp. 2049--2058, 2015.

\bibitem{avsravaliable}
S.~Dupont and J.~Luettin, ``{Audio-visual speech modeling for continuous speech
  recognition},'' \emph{IEEE Transactions on Multimedia}, vol.~2, no.~3, pp.
  141--151, 2000.

\bibitem{avaliablemulti}
S.~Thermos and G.~Potamianos, ``{Audio-visual speech activity detection in a
  two-speaker scenario incorporating depth information from a profile or
  frontal view},'' in \emph{2016 IEEE Spoken Language Technology Workshop
  (SLT)}, 2016, pp. 579--584.

\bibitem{multiview}
C.~Xu, Y.~Wang, T.~Tan, and L.~Quan, ``{Depth vs. intensity: which is more
  important for face recognition?}'' in \emph{Proceedings of the 17th
  International Conference on Pattern Recognition, 2004. ICPR 2004.}, vol.~1,
  2004, pp. 342--345 Vol.1.

\bibitem{liu2018automatic}
L.~Liu, G.~Feng, and D.~Beautemps, ``Automatic temporal segmentation of hand
  movements for hand positions recognition in french cued speech,'' in
  \emph{2018 IEEE ICASSP}.\hskip 1em plus 0.5em minus 0.4em\relax IEEE, 2018,
  p. 3061–3065.

\bibitem{liu2018visual}
L.~Liu, T.~Hueber, G.~Feng, and D.~Beautemps, ``Visual recognition of
  continuous cued speech using a tandem cnn-hmm approach.'' in
  \emph{Interspeech}, 2018, p. 2643–2647.

\bibitem{AVLetter}
I.~Matthews, T.~Cootes, S.~Cox, R.~Harvey, and J.~A. Bangham, ``{Lipreading
  using shape, shading and scale},'' in \emph{AVSP'98 International Conference
  on Auditory-Visual Speech Processing}, 1998.

\bibitem{AVLetter-2}
S.~Cox, R.~Harvey, and Y.~Lan, ``{The Challenge of Multispeaker Lip-Reading},''
  in \emph{Proc of International Conference on Auditory-visual Speech
  Processing}, 2008.

\bibitem{DAVID}
C.~C. Chibelushi, F.~Deravi, and J.~Mason, ``{A review of speech-based bimodal
  recognition},'' \emph{IEEE Transactions on Multimedia}, vol.~4, no.~1, pp.
  23--37, 2002.

\bibitem{GRID}
M.~Cooke, J.~Barker, S.~Cunningham, and X.~Shao, ``{An audio-visual corpus for
  speech perception and automatic speech recognition},'' \emph{The Journal of
  the Acoustical Society of America}, vol. 120, no. 5 Pt 1, p. 2421—2424,
  November 2006.

\bibitem{cavsr}
Y.~Xu, L.~Du, G.~Li, P.~Wu, and X.~Zhang, ``{Chinese audiovisual bimodal speech
  database CAVSR1.0},'' in \emph{Proc. Int. Symp. Chin. Spoken Lang. Process.},
  2000, pp. 98--101.

\bibitem{hit2}
X.~Lin, H.~Yao, and Hong, ``{A sentence-level lip-reading-based corpus and its
  slice and dice algorithm},'' \emph{Computer Engineering and Applications},
  vol.~41, no.~3, pp. 174--177, 2005.

\bibitem{liu2017automatic}
L.~Liu, G.~Feng, and D.~Beautemps, ``Automatic dynamic template tracking of
  inner lips based on clnf,'' in \emph{2017 ICASSP}.\hskip 1em plus 0.5em minus
  0.4em\relax IEEE, 2017, p. 5130–5134.

\bibitem{hit}
X.~Lin, H.~Yao, X.~Hong, and Q.~Wang, ``{HIT-AVDB-II: A new multi-view and
  extreme feature cases contained audio-visual database for biometrics},'' in
  \emph{11th Joint International Conference on Information Sciences}.\hskip 1em
  plus 0.5em minus 0.4em\relax Atlantis Press, 2008, pp. 357--363.

\bibitem{LRW-1000}
S.~Yang, Y.~Zhang, D.~Feng, M.~Yang, C.~Wang, J.~Xiao, K.~Long, S.~Shan, and
  X.~Chen, ``{LRW-1000: A Naturally-Distributed Large-Scale Benchmark for Lip
  Reading in the Wild},'' in \emph{2019 14th IEEE International Conference on
  Automatic Face \& Gesture Recognition (FG 2019)}, 2019, pp. 1--8.

\bibitem{SITW}
M.~Mclaren, L.~Ferrer, D.~Castan, and A.~Lawson, ``{The Speakers in the Wild
  (SITW) Speaker Recognition Database},'' in \emph{Interspeech 2016}, 2016.

\bibitem{voxceleb}
A.~Nagrani, J.~S. Chung, and A.~Zisserman, ``{VoxCeleb: A Large-Scale Speaker
  Identification Dataset},'' in \emph{Interspeech 2017}.\hskip 1em plus 0.5em
  minus 0.4em\relax {ISCA}, aug 2017.

\bibitem{zhang2022webuav}
C.~Zhang, G.~Huang, L.~Liu, S.~Huang, Y.~Yang, X.~Wan, S.~Ge, and D.~Tao,
  ``Webuav-3 m: A benchmark for unveiling the power of million-scale deep uav
  tracking,'' \emph{IEEE TPAMI}, 2022.

\bibitem{CUAVE}
E.~Patterson, S.~Gurbuz, Z.~Tufekci, and J.~Gowdy, ``{CUAVE: A new audio-visual
  database for multimodal human-computer interface research},'' in \emph{2002
  IEEE International Conference on Acoustics, Speech, and Signal Processing},
  vol.~2, 2002, pp. II--2017--II--2020.

\bibitem{GAMVA}
S.~Isobe, R.~Hirose, T.~Nishiwaki, T.~Hattori, S.~Tamura, Y.~Gotoh, and
  M.~Nose, ``{GAMVA: A Japanese Audio-Visual Multi-Angle Speech Corpus},'' in
  \emph{2021 24th Conference of the Oriental COCOSDA International Committee
  for the Co-ordination and Standardisation of Speech Databases and Assessment
  Techniques (O-COCOSDA)}, 2021, pp. 134--139.

\bibitem{wapusk20}
A.~Vorwerk, X.~Wang, D.~Kolossa, S.~Zeiler, and R.~Orglmeister, ``{WAPUSK20-A
  Database for Robust Audiovisual Speech Recognition.}'' in \emph{LREC}.\hskip
  1em plus 0.5em minus 0.4em\relax Citeseer, 2010.

\bibitem{bavcd}
G.~Galatas, G.~Potamianos, D.~Kosmopoulos, C.~McMurrough, and F.~Makedon,
  ``{Bilingual corpus for AVASR using multiple sensors and depth
  information},'' in \emph{Auditory-Visual Speech Processing 2011}, 2011.

\bibitem{rgbd}
N.~Ahmed, ``{RGB-D dynamic facial dataset capture for visual speech
  recognition},'' in \emph{2019 International Conference on Image and Video
  Processing, and Artificial Intelligence}, vol. 11321.\hskip 1em plus 0.5em
  minus 0.4em\relax SPIE, 2019, pp. 42--46.

\bibitem{wav2vec}
A.~Baevski, Y.~Zhou, A.~Mohamed, and M.~Auli, ``{wav2vec 2.0: A framework for
  self-supervised learning of speech representations},'' \emph{Advances in
  neural information processing systems}, vol.~33, pp. 12\,449--12\,460, 2020.

\bibitem{resnet}
K.~He, X.~Zhang, S.~Ren, and J.~Sun, ``{Deep residual learning for image
  recognition},'' in \emph{Proceedings of the IEEE conference on computer
  vision and pattern recognition}, 2016, pp. 770--778.

\bibitem{dlib}
D.~E. King, ``{Dlib-ml: A machine learning toolkit},'' \emph{The Journal of
  Machine Learning Research}, vol.~10, pp. 1755--1758, 2009.

\bibitem{aishell-2}
J.~Du, X.~Na, X.~Liu, and H.~Bu, ``{Aishell-2: Transforming mandarin asr
  research into industrial scale},'' \emph{arXiv preprint arXiv:1808.10583},
  2018.

\bibitem{AISHELL-1}
H.~Bu, J.~Du, X.~Na, B.~Wu, and H.~Zheng, ``{AISHELL-1: An open-source Mandarin
  speech corpus and a speech recognition baseline},'' in \emph{2017 20th
  O-COCOSDA}, 2017, pp. 1--5.

\bibitem{BILSTM}
A.~Graves, A.-r. Mohamed, and G.~Hinton, ``{Speech recognition with deep
  recurrent neural networks},'' in \emph{2013 IEEE ICASSP}.\hskip 1em plus
  0.5em minus 0.4em\relax Ieee, 2013, pp. 6645--6649.

\bibitem{ctc}
A.~Graves, S.~Fern{\'a}ndez, F.~Gomez, and J.~Schmidhuber, ``{Connectionist
  temporal classification: labelling unsegmented sequence data with recurrent
  neural networks},'' in \emph{Proceedings of the 23rd international conference
  on Machine learning}, 2006, pp. 369--376.

\end{thebibliography}

\end{document}